 \documentclass[aps,prd,reprint,twocolumn,showpacs,superscriptaddress,%
nobibnotes,nofootinbib]{revtex4-1} 

\def\mysections#1{{\bf #1.} }

\usepackage[dvips]{graphicx}
\usepackage{epsfig,amsmath,amssymb,verbatim,mathrsfs,array,layout,textcomp,latexsym, slashed,graphicx,bbm}

\usepackage[normalem]{ulem}
\usepackage{appendix}

\usepackage{rotating}
\usepackage{hyperref}
\usepackage[usenames,dvipsnames]{color}

\usepackage{soul}

\newcommand{\be}{\begin{equation}}
\newcommand{\ee}{\end{equation}}
\newcommand{\bea}{\begin{eqnarray}}
\newcommand{\eea}{\end{eqnarray}}

\definecolor{gbcolor}{rgb}{.8,.3,.1}
\definecolor{gbcolor2}{rgb}{.8,.1,.7}

\newcommand{\AR}[1]{\color{black} #1 \color{black}}

\DeclareMathOperator{\sinc}{sinc}

\def\beq{\begin{equation}}
\def\eeq{\end{equation}}

\begin{document}

\begin{flushleft}                          
\footnotesize DESY 18-128 \
\end{flushleft} 

\begin{flushleft}                          
\end{flushleft}

\title{Measuring the Boiling Point of the Vacuum of Quantum Electrodynamics}

\author{Anthony Hartin}\email{anthony.hartin@desy.de}
\affiliation{University College London, Gower Street, London WC1E 6BT, United Kingdom}
\author{Andreas Ringwald}\email{andreas.ringwald@desy.de}
\affiliation{DESY, Notkestrasse 85, 22607 Hamburg, Germany}
\author{Natalia Tapia}\email{natalia.tapiaa@usach.cl}
\affiliation{Departamento de Fisica, Universidad de Santiago de Chile, Casilla 307, Santiago, Chile}



\begin{abstract}
It is a long-standing non-trivial prediction of quantum electrodynamics that its vacuum is unstable in the background of a static, spatially 
uniform electric field and, in principle, sparks with spontaneous emission of electron-positron pairs. 
However, an experimental verification of this prediction seems out of reach because a sizeable rate for spontaneous pair production requires 
an extraordinarily strong electric field strength $|\mathbf E|$ of order 
the Schwinger critical field, ${\rm E}_c=m_e^2/e\simeq 1.3\times 10^{18}\ {\rm V/m}$, where $m_e$ is the 
electron mass and $e$ is its charge. 
Here, we show that the measurement of the rate of pair production due to the decays of
high-energy bremsstrahlung photons in a high-intensity laser field allows for the experimental determination of the
Schwinger critical field and thus the boiling point of the vacuum of quantum electrodynamics.
\end{abstract}

\maketitle

\section{Introduction}

Quantum Electrodynamics (QED) is one of the most successful theories in physics. 
Its predictions for observables accessible by an ordinary perturbative expansion in the electromagnetic coupling
$e$, such as for example for the anomalous magnetic moment of the electron, have been verified experimentally 
to a very high accuracy. 

There are, however, also observables which are inaccessible by ordinary perturbation theory and whose  
prediction lacks an experimental verification. 
Among them, the most famous is the rate  (per unit volume $V$) of spontaneous electron-positron pair production (SPP)  
in a strong static electric field
$\bf E$~\cite{Sauter:1931,Heisenberg:1936qt,Schwinger:1951nm} , 
\begin{eqnarray}
\label{schwinger-rate}
\frac{\Gamma_{\text{SPP}}}{V} 
=\frac{m^{4}_e}{(2\pi)^3}
\left(\frac{{\bf{|E|}}}{\rm{E}_{c}}\right)^{2}\sum^{\infty}_{n=1}
\frac{1}{n^2}\exp\left(-n\pi\frac{\rm{E}_{c}}{{\bf{|E|}}}\right)\,,
\end{eqnarray}
where  
\begin{equation}
{\rm E}_c \equiv 
\frac{m_e^2}{e}
\simeq 1.3\times 10^{18}\ {\rm V/m}
\label{schwinger-crit}
\end{equation}
is the so-called Schwinger critical field. Clearly, this rate is non-perturbative in $e$, 
\begin{equation}
\Gamma_{\rm SPP} \propto\exp\left(-\pi\frac{m_e^2}{e{\bf{|E|}}}\right)\,,
\end{equation} 
as typical 
for a process which can occur, for  ${\bf |E|}\lesssim {\rm E}_c$, only via quantum tunnelling. 
This so-called Schwinger effect and its analogues have been suggested to play a role in many 
problems of phenomenological and cosmological interest, ranging from black hole quantum 
evaporation~\cite{Hawking:1975sw,Damour:1976jd,Gibbons:1978pt,Gavrilov:1996pz}
to particle production in hadronic 
collisions~\cite{Casher:1979wy,Andersson:1983ia,Biro:1984cf} 
and in the early 
universe~\cite{Parker:1969au,Birrell:1982,Tangarife:2017rgl}, to 
mention only a few. 
Unfortunately, there is no practical way to produce a static electric field 
of this strength in the foreseeable future\footnote{One possibility considered was the field at the crossing of two intense laser 
beams \cite{Bunkin:1970,Brezin:1970,Popov:1971,Popov:1974}. However, the required laser peak power is in the hundreds of exawatt range  (for a laser operating in the optical range, focussed 
to the diffraction limit)  \cite{Ringwald:2001ib}  and thus still far beyond the present technology. However, the pair production can be
strongly enhanced if one superimposes the intense field at the laser focus with a further weak high frequency electromagnetic field \cite{Schutzhold:2008pz,Dunne:2009gi}, paving the way to a possible detectibility at the Extreme Light Infrastructure ELI \cite{Narozhny:2015vsb,Turcu:2016dxm}.}. Therefore, a direct laboratory test 
 of prediction  \eqref{schwinger-rate}  
seems utopic. 

\begin{figure}
\begin{center}
\includegraphics[width=\columnwidth]{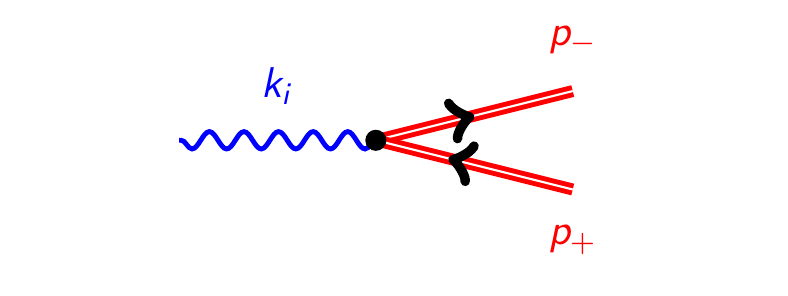}
\vspace{-5ex}
\caption{Leading order Furry picture \cite{Furry:1951zz} Feynman diagram for OPPP. The double line pointing forward (backward) in time
represents an electron (a positron) in the background of the electromagnetic field of the laser. 
\label{fig:diagram_oppp}
}
\end{center}
\end{figure}

As an alternative to spontaneous pair production in a strong static electric field, we consider here laser-assisted 
one photon pair production (OPPP) -- the decay of a high energy photon in the overlap with an intense optical laser beam into an electron-positron pair, cf. Fig. \ref{fig:diagram_oppp}. 
This process is kinematically possible because the electron-positron pair can pick up momentum from the laser photons. 
Already in the 1960`s, when first lasers where developed, this process has been identified  as an opportunity to study the transition from stimulated to spontaneous pair production in an external electromagnetic field \cite{Reiss:1962,Narozhnyi:1965}.\footnote{The OPPP 
process continues to attract much modern interest with new theoretical approaches \cite{Lv:2018qxy}, analyses which take into account 
real interacting laser pulses \cite{Jansen:2015idl,Otto:2016fdo,DiPiazza:2016maj,DiPiazza:2015xva} and experimental schemes to realise the 
process \cite{Lobet:2017vug}.} 
We will show in this paper that it offers a timely way to probe the so far elusive boiling of the vacuum of 
QED and to determine the Schwinger critical field experimentally.

The paper is organised as follows. We first examine the transition rate of the OPPP process as it varies with the laser field intensity and the photon recoil parameter (cf. Sec. \ref{sec:OPPP}). The rate will be shown to be well described by asymptotic expressions, which depend on the Schwinger critical field in a simple way, for both low high laser intensities. Next we will consider in Sec. \ref{sec:BPPP} the effect of generating high energy photons via bremsstrahlung from a foil on the asymptotic features of the rate. Finally we consider 
in Sec. \ref{sec:experiment} real experimental parameters and the effect of the finite duration of the laser pulse and the variability of the laser intensity throughout the interaction region on the determined value of the Schwinger critical field. We conclude in Sec. \ref{sec:conclusions}.

\begin{figure}
\begin{center}
\includegraphics[width=0.9\columnwidth]{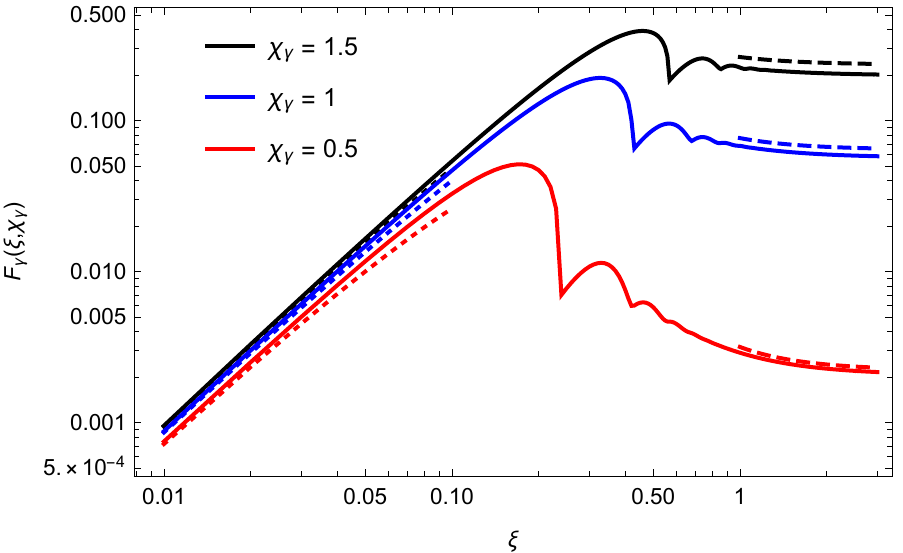}
\vspace{-1ex}
\caption{The dimensionless function $F_\gamma(\xi,\chi_\gamma )$, Eq. \eqref{eq:Foppp}, describing the 
probability of laser-assisted OPPP, as a function of the laser intensity parameter $\xi$, for different 
values of the photon recoil parameter $\chi_\gamma$ (solid lines). The dotted (dashed) line shows the 
analytic result valid at small (large) values of the intensity parameter, Eq. \eqref{pair_prod_circularly_pol_laser_low_intensity}
(Eq. \eqref{pair_prod_circularly_pol_laser_large_intensity}). 
\label{fig:log_log_decay_rate}
}
\end{center}
\end{figure}

\section{One photon pair production}
\label{sec:OPPP}

To leading order in Furry picture \cite{Furry:1951zz} perturbation theory  (for a recent review, see Ref. \cite{Hartin:2018egj}), the rate of 
laser-assisted OPPP 
can be written in the form
\begin{equation}
\Gamma_{\text{OPPP}}=  \frac{\alpha m_e^2}{4\,\omega_\text{i}}  \, 
F_\gamma(\xi , \chi_\gamma ) \,,
\label{gamma_oppp}
\end{equation}
where $\alpha = e^2/(4\pi)$ is the fine structure constant, $k_i =( \omega_\text{i},{\mathbf{k}}_\text{i})$, with $\omega_i^2={\mathbf{k}}_\text{i}^2$, is the four-momentum 
of the initial state photon, and $\xi$ and $\chi_\gamma$ are the laser intensity parameter and the photon recoil parameter, respectively,
\begin{widetext} 
\begin{eqnarray}
\xi\equiv \frac{e\left|\mathbf{E} \right|}{\omega m_e}
=\frac{m_e}{\omega}\frac{\left|\mathbf{E} \right|}{{\rm E}_c}\,, \quad \quad \quad \quad
 \chi_\gamma\equiv \frac{k\cdot k_i}{m_e^2} 
\,  \xi 
= (1+\cos\theta )\, \frac{\omega_i}{m_e}\,  \frac{\left|\mathbf{E} \right|}{{\rm E}_c}
\,,
\label{photon_recoil}
\end{eqnarray}
in terms of the electric field $\left|\mathbf{E} \right|$ of the laser beam, its frequency $\omega$, and its angle $\theta$ with 
respect to the direction of the incident photon.  The dimensionless function $F_\gamma(\xi ,\chi_\gamma)$, for 
the idealized case that the electromagnetic field of the laser beam can be described as a circularly polarized infinite plane 
wave (IPW),\footnote{The general problem of assisted pair production from counter propagating laser/photon pulses was considered, primarily from a theoretical standpoint, in Ref. \cite{Fedotov:2013uja}. Three theoretical approximations were considered, the delta pulse method involving the overlap of in and out states applicable for $\chi_\gamma\gg 1$, the locally constant field approximation for $\xi\gg 1$, and perturbation theory for $\xi\ll 1$. Instead, for the scheme proposed in this paper, with high energy photons produced from foil bremsstrahlung, the initial states vary widely across energy and spatial ranges. Preliminary analysis shows that the IPW approximation, using local values of strong field parameters, are a suitably accurate description for the real experiments being envisaged.} 
is given by a sum over the effective number of laser photons $n$ absorbed by the electron-positron pair \cite{Narozhnyi:1965},  
\begin{eqnarray}
\label{eq:Foppp}
F_\gamma (\xi , \chi_\gamma ) =
\sum\limits_{n>n_\text{o}}^{\infty}\int_1^{v_n} \frac{\text{d}v}{v\sqrt{v(v-1)}} \left[ 2\,J_{n}^2(z_v)+\xi^2(2v-1)\left( J_{n\text{+1}}^2(z_v)+J_{n\text{-1}}^2(z_v)-2 J_{n}^2(z_v)\right) \right] ,
\end{eqnarray}
with Bessel functions $J_n$ and 
\begin{equation} 
n_0 \equiv 
\frac{2\xi\left(1+ \xi^2\right)}{\chi_\gamma} ,  \quad\quad
z_v\equiv \frac{4 \xi^2\sqrt{1+\xi^2}}{\chi_\gamma}\left[ v\left( v_n-v\right)\right]^{1/2}, \quad\quad
v_n\equiv \frac{\chi_\gamma\,n}{2\xi(1+\xi^2)}. 
\end{equation}
\end{widetext}

In Fig. \ref{fig:log_log_decay_rate}, we display $F_\gamma(\xi ,\chi_\gamma )$  
as a function of $\xi$, for three values of $\chi_\gamma$. 
Clearly, at low laser intensities, $\xi\ll 1$, laser-assisted OPPP appears to proceed perturbatively, 
$F_\gamma\propto \xi^2 \propto \alpha$, as expected from the necessity to absorb at least one laser photon
to allow for photon decay kinematically.  
In fact, expanding $F_\gamma$ for small $\xi$ yields 
\begin{equation}
\label{pair_prod_circularly_pol_laser_low_intensity}
F_\gamma (\xi , \chi_\gamma ) =  
2\,\xi^2 \left[ \log \left( \frac{2\chi_\gamma}{\xi}\right) -1 \right] + {\mathcal O} 
\left( \xi^3 \log\,{\xi} \right) .
\end{equation}
This behaviour reproduces the full result for laser-assisted OPPP up to values of $\xi \sim 0.1$, cf. Fig. \ref{fig:log_log_decay_rate}.  
As the laser intensity $\xi$ increases, the threshold number of absorbed photons $n_0$ to produce an electron-positron pair 
increases, and more and more terms in the summation over the number
of absorbed laser photons in Eq. \eqref{eq:Foppp} drop out of the probability, resulting in the appearance
of less and less pronounced maxima in $F_\gamma$, see Fig. \ref{fig:log_log_decay_rate}. 
At large $\xi$, finally, the probability of laser-assisted OPPP approaches a finite value, the 
latter growing with increasing $\chi_\gamma$. Indeed, for 
$\xi\gtrsim 1/\sqrt{\chi_\gamma}\gg 1$, $F_\gamma$ behaves as \cite{Ritus:1985}
\begin{eqnarray}
\label{pair_prod_circularly_pol_laser_large_intensity}
F_\gamma (\xi , \chi_\gamma ) = 
\frac{3}{4} 
\sqrt{\frac{3}{2}}\,
\chi_\gamma
\,
{\rm e}^{\left[ 
-\frac{8}{3\chi_\gamma} 
\left( 1-\frac{1}{15}\xi^{-2}+\mathcal O(\xi^{-4})\right)
\right]} .
\end{eqnarray}
This behaviour applies to a very good accuracy already for $\xi\gtrsim 1$ and $\chi_\gamma\lesssim 1$, 
cf. Fig.  \ref{fig:log_log_decay_rate}. 
Importantly, we infer from Eq. \eqref{pair_prod_circularly_pol_laser_large_intensity} that the 
asymptotic value of $F_\gamma$ is non-perturbative in the electromagnetic coupling $e$ and that the rate of 
laser-assisted OPPP asymptotes to  
\begin{widetext}
\begin{equation}
\label{eq:SPP_OPPP}
\Gamma_{\text{OPPP}} 
\to  \frac{3}{16} 
\sqrt{\frac{3}{2}}\,\alpha\, m_e\, 
 (1+\cos\theta )\, \frac{}{}\,  \frac{\left|\mathbf{E} \right|}{{\rm E}_c} 
\exp \left[ 
-\frac{8}{3} \frac{1}{1+\cos\theta}\frac{m_e}{\omega_i}\frac{{\rm E}_c}{\left|\mathbf{E} \right|}
\right] \,,
\end{equation}
\end{widetext}
ressembling the rate \eqref{schwinger-rate} of SPP in a constant electric field\footnote{The leading term in the exponent in Eq. \ref{eq:SPP_OPPP}
is independent of the laser polarisation, while the pre-factor depends on it  \cite{Ritus:1985}}. 
This has to be expected, since large intensity parameter, $\xi\gg 1$, corresponds to a quasi-static electric field of the laser, 
$\omega \ll e\,\left|\mathbf{E} \right|/m_e$, cf. Eq.  \eqref{photon_recoil}. 
However, in contrast to SPP,  in laser-assisted OPPP the produced electron-positron pair, in its rest frame, experiences an electric field enhanced by the 
relativistic boost factor $\omega_i/m_e$. This enhanced electric field is of order the Schwinger critical 
value ${\rm E}_c$, if  the photon recoil parameter is $\chi_\gamma\sim 1$, cf. Eq. \eqref{photon_recoil}. 
Hence, the Schwinger critical field -- the boiling point of the QED vacuum -- can be determined in principle experimentally from the
measurement of the rate of laser-assisted OPPP at $\xi \gtrsim 1/\sqrt{\chi_\gamma}\gg 1$.
Next, we consider the effect of enhancing the OPPP rate with the use of high energy 
bremsstrahlung photons.

\section{Bremsstrahlung photon pair production}
\label{sec:BPPP}

\begin{figure}
\begin{center}
\includegraphics[width=\columnwidth]{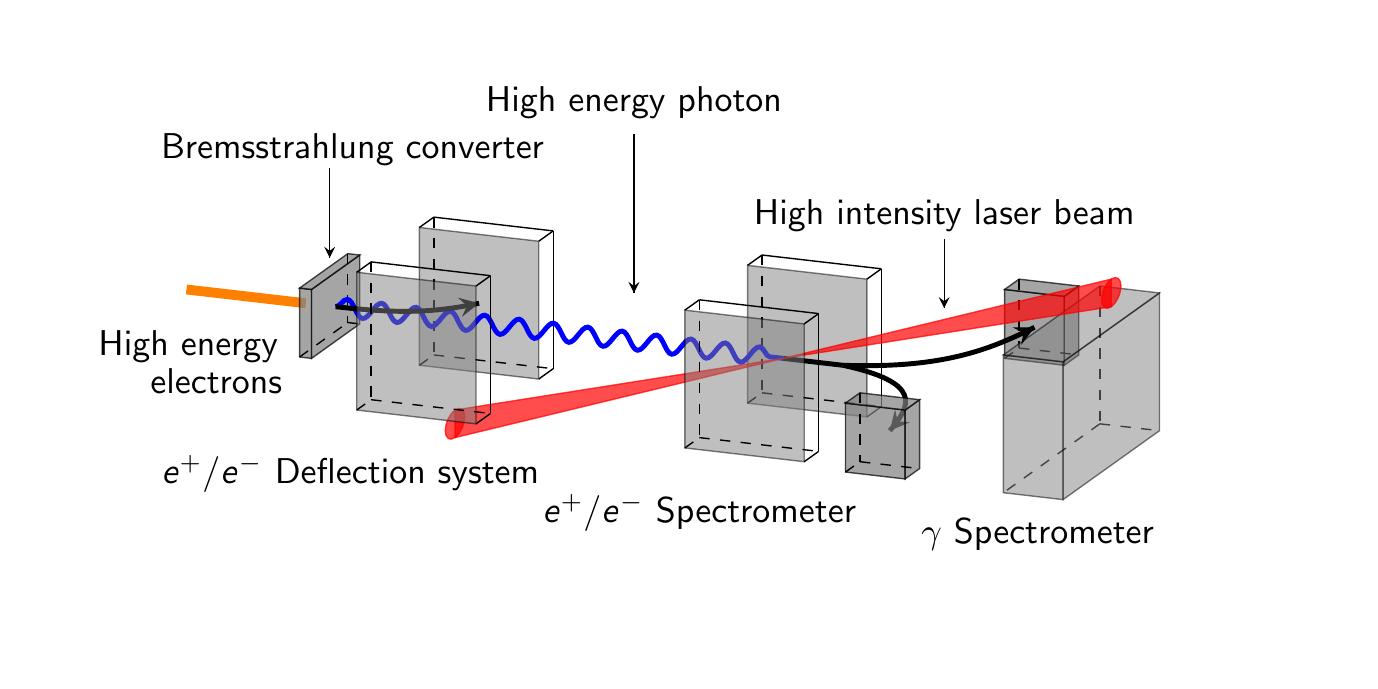}
\vspace{-8ex}
\caption{Sketch of an experiment to produce high energy photons by bremsstrahlung conversion in a high-$Z$ thin target 
and to cross them with a laser beam to let them decay into electron-positron pairs. Switching off the laser allows for a determination
of the bremsstrahlung spectrum. Removing the target allows in addition for the study of HICS, followed by
OPPP, and of the one-step trident process. 
\label{fig:experiment_sketch}
}
\end{center}
\end{figure}

Note, that for a laser of frequency $\omega=1\,\text{eV}$, focussed to an intensity $I$, corresponding to\footnote{This relation assumes that the intensity is given by the modulus of the Pointing vector, i.e. $I=\left|\mathbf{E} \right|^2$ for a plane wave.} 
\AR{\begin{equation}
\label{xi_intensity}
\xi = 2.370 \left( \frac{I}{10^{19}\,\text{W}/\text{cm}^2}\right)^{1/2} \left( \frac{1\,\text{eV}}{\omega}\right) \,,
\end{equation}}
the condition $\xi \gtrsim 1/\sqrt{\chi_\gamma}$ leads to a lower bound on the energy of the high energy photon, 
\AR{$\omega_i \gtrsim {19.6\,\text{GeV}}\left( \frac{1\,\text{eV}}{\omega}\right)\frac{\left( {2.37}/{\xi}\right)^{3}}{(1+\cos\theta)}$.} 
Unfortunately, there are no mono-energetic photon beams with energies in the $\mathcal O(10)$\,GeV range available. 
On the other hand, there are $\mathcal O(10)$\,GeV electron beams, notably the ones exploited by 
X-ray free electron lasers, such as LCLS \cite{Arthur:2002ap} in Stanford or the European XFEL \cite{Altarelli:2006zza} in Hamburg. 
Such an electron beam can be sent to a high-$Z$ target in which it is converted by bremsstrahlung into a collimated high energy photon beam, which can then be crossed with a high-intensity laser beam, cf. Fig. \ref{fig:experiment_sketch}. 
Such an experiment to study laser-assisted bremsstrahlung photon pair production (BPPP) has been envisaged long time ago in Ref. \cite{Reiss:1971wf} and more recently  
discussed in Refs.   \cite{Gemini_proposal:2010,Turcu:2016dxm,Blackburn:2018ghi}. Here, we show 
that even after integration over the bremsstrahlung spectrum, the Schwinger critical field can be 
determined experimentally from the measurement of the total rate of electron-positron pair production at large laser intensity.

Given the energy spectrum $\text{d}N_\gamma/\text{d}\omega_i$ of photons generated by an electron impinging on the foil, 
the rate of laser-assisted BPPP  
is given by 
\begin{eqnarray}
\nonumber
\Gamma_{\text{BPPP}} 
&= & \frac{\alpha\, m_e^2}{4}  \int_0^{E_e} \frac{\text{d}\omega_i}{\omega_\text{i}} \,\frac{\text{d}N_\gamma}{\text{d}\omega_i} \,
F_\gamma (\xi , \chi_\gamma (\omega_i)) 
\\ 
\label{electron_initated_laser_assisted_OPPP}
&= & \frac{\alpha\, m_e^2}{4}  
\frac{\chi_e}{E_e}\int_0^{\chi_e} \frac{\text{d}\chi_\gamma}{\chi_\gamma} \,\frac{\text{d}N_\gamma}{\text{d}\chi_\gamma} \,
F_\gamma (\xi , \chi_\gamma) \,,
\end{eqnarray} 
where $E_e$ is the energy of the incident electrons and $\chi_e \equiv k\cdot k_e\,\xi/ m_e^2=(1+\cos\theta )\omega E_e \xi/m_e^2$
is the electron recoil parameter. 

\begin{figure}
\begin{center} 
\includegraphics[width=0.99\columnwidth]{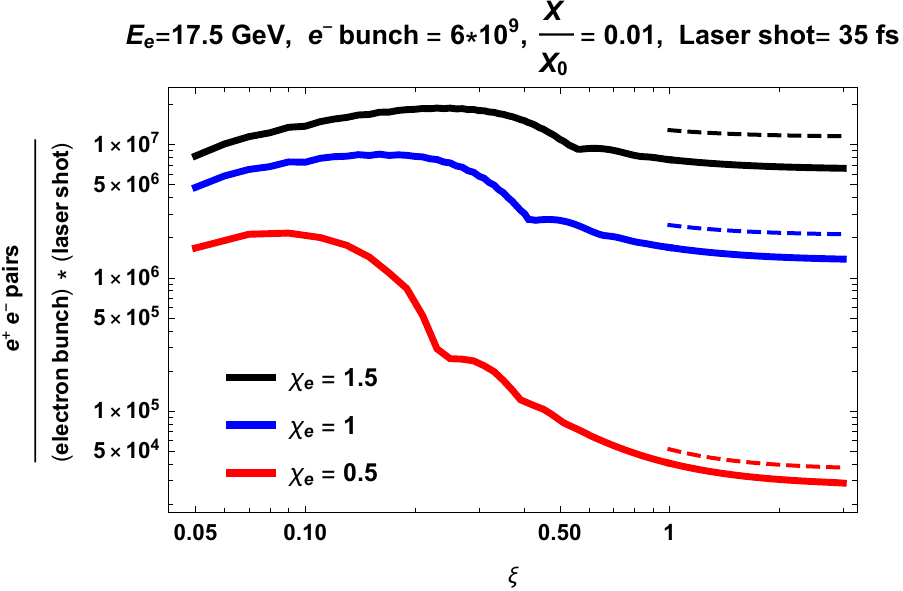}
\vspace{-2ex}
\caption{Number of $e^+ e^-$ pairs produced per electron bunch ($6\times 10^9$ electrons of energy $E_e=17.5\,\text{GeV}$) impinging on the converter target (thickness $X/X_0=0.01$) and per laser shot (duration $35\,\text{fs}$)
crossed with the bremsstrahlung photons, as a function of the laser intensity parameter $\xi$, for different values of $\chi_e$. 
The dashed  line shows the 
analytic prediction resulting from \eqref{pair_prod_circularly_pol_laser_large_intensity_thin_target_brems}, 
valid at $\xi\gtrsim 1/\sqrt{\chi_e}\gg 1$. 
\label{fig:log_log_decay_rate_thin_target_brems}
}
\end{center}
\end{figure}

\begin{figure}
\begin{center} 
\includegraphics[width=0.9\columnwidth]{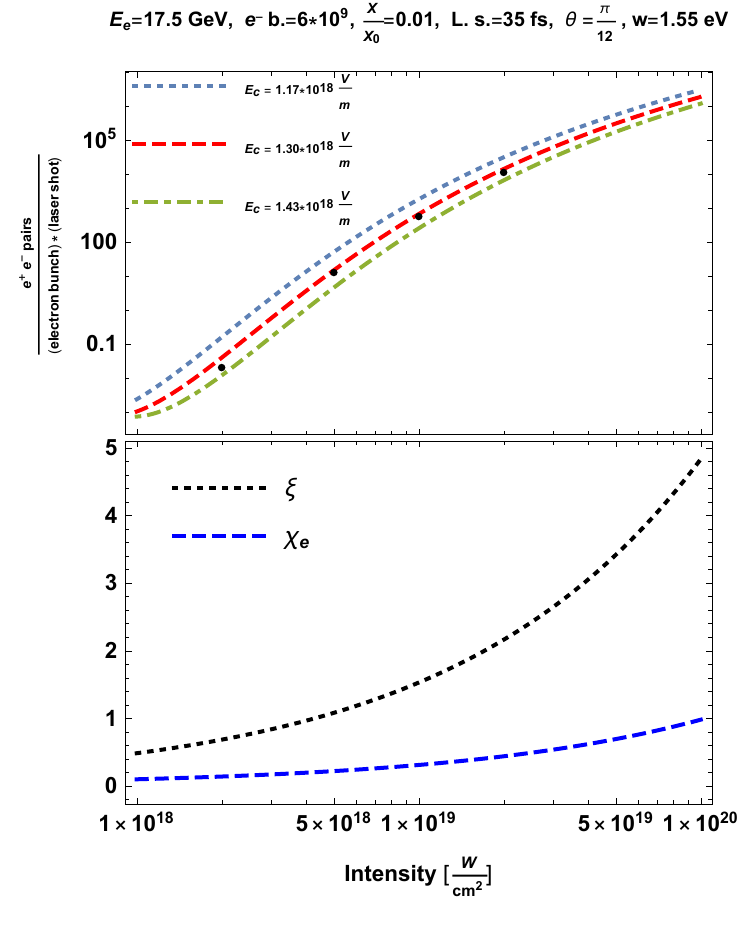}
\vspace{-4ex}
\caption{\AR{{\em Top panel:} Total number of $e^+ e^-$ pairs produced per electron bunch ($6\times 10^9$ electrons of energy $E_e=17.5\,\text{GeV}$) impinging on the bremsstrahlung target (thickness $X/X_0=0.01$) and per laser shot (duration $35\,\text{fs}$, laser frequency $\omega = 1.55\,\text{eV}$, corresponding to a laser wavelength of 800 nm) 
crossed with the bremsstrahlung photons at an angle of $\theta = \pi/12$, as a function of the laser intensity. The black dots exploit the 
full numerical result for specific values of the intensity. 
The dashed line shows the
analytic prediction resulting from \eqref{pair_prod_circularly_pol_laser_large_intensity_thin_target_brems},
exploiting the relations \eqref{xi_intensity} and \eqref{chie_intensity}. The dotted (dot-dashed) line shows the same analytic prediction,
but for the case where the value of the Schwinger critical field ${\rm E}_c$ deviates by a multiplicative factor of $\kappa =0.9$ ($\kappa = 1.1$) from its nominal value 
\eqref{schwinger-crit}. 
{\em Bottom panel:} The laser intensity parameter $\xi$ (dotted) and the electron recoil parameter (dashed), as a function of the intensity, cf. Eqs. \eqref{xi_intensity} and \eqref{chie_intensity}.
\label{fig:log_log_decay_rate_thin_target_brems_intensity}
}
}
\end{center}
\end{figure}

For a target of thickness $X \ll X_0$, where $X_0$ is the radiation length, the bremsstrahlung spectrum can be approximated by \cite{Tsai:1973py}
\begin{equation}
\label{brems_spectrum_thin_target}
\omega_i\frac{\text{d}N_\gamma}{\text{d}\omega_i} \approx
\left[ \frac{4}{3} - \frac{4}{3}\left( \frac{\omega_i}{E_e}\right) +\left(\frac{\omega_i}{E_e}\right)^2\right] \frac{X}{X_0} \,,
\end{equation} 
if one assumes complete screening.\footnote{We have checked via Monte Carlo simulations with GEANT \cite{Agostinelli:2002hh} that \eqref{brems_spectrum_thin_target} is valid in the parameter range we use it in e.g. Figs.  \ref{fig:log_log_decay_rate_thin_target_brems}, \ref{fig:log_log_decay_rate_thin_target_brems_intensity} and \ref{fig:log_log_decay_rate_thin_target_brems_intensity_Ee}. For the interpretation of the experiment itself one does not have to rely on a theoretical prediction, since the bremsstrahlung spectrum can be measured by switching off the laser, cf. Fig. \ref{fig:experiment_sketch}.} 
This results, at high laser intensities, $\xi\gtrsim 1/\sqrt{\chi_e}\gg 1$, in the non-perturbative, $e^{-8/(3\chi_e)}$ dependence
of the laser-assisted BPPP rate, 
\begin{eqnarray}
\label{pair_prod_circularly_pol_laser_large_intensity_thin_target_brems}
\Gamma_{\text{BPPP}} 
\to 
 \frac{\alpha\, m_e^2}{E_e} \frac{9}{128} \sqrt{\frac{3}{2}}\, \chi_e^2\, e^{-\frac{8}{3 \text{$\chi_e$}} 
\left( 1 - \frac{1}{15 \xi^{2}} \right)}   \frac{X}{X_0} \,,
\end{eqnarray}
ressembling the behavior of the laser-assisted OPPP rate, Eqs. \eqref{gamma_oppp} and \eqref{pair_prod_circularly_pol_laser_large_intensity},  if one replaces in the latter expression $\chi_\gamma$ by $\chi_e$. 
Therefore, the Schwinger critical field can be inferred from the asymptotic behavior of laser-assisted BPPP for high laser intensities,
\begin{widetext}
\AR{\begin{equation}
\label{eq:SPP_BPPP}
\Gamma_{\text{BPPP}} 
\to  
 \frac{9}{128} \sqrt{\frac{3}{2}}\, \alpha\, E_e\,(1+\cos\theta )^2
\left( \frac{\left|\mathbf{E} \right|}{{\rm E}_c}\right)^2
\exp \left[ 
-\frac{8}{3} \frac{1}{1+\cos\theta}\frac{m_e}{E_e}\frac{{\rm E}_c}{\left|\mathbf{E} \right|}
\right] 
 \frac{X}{X_0} \,.
\end{equation}}
\end{widetext}

\section{Experimental considerations}
\label{sec:experiment} 

\subsection{Expected sensitivity to critical field}

\begin{figure}
\begin{center} 
\includegraphics[width=0.95\columnwidth]{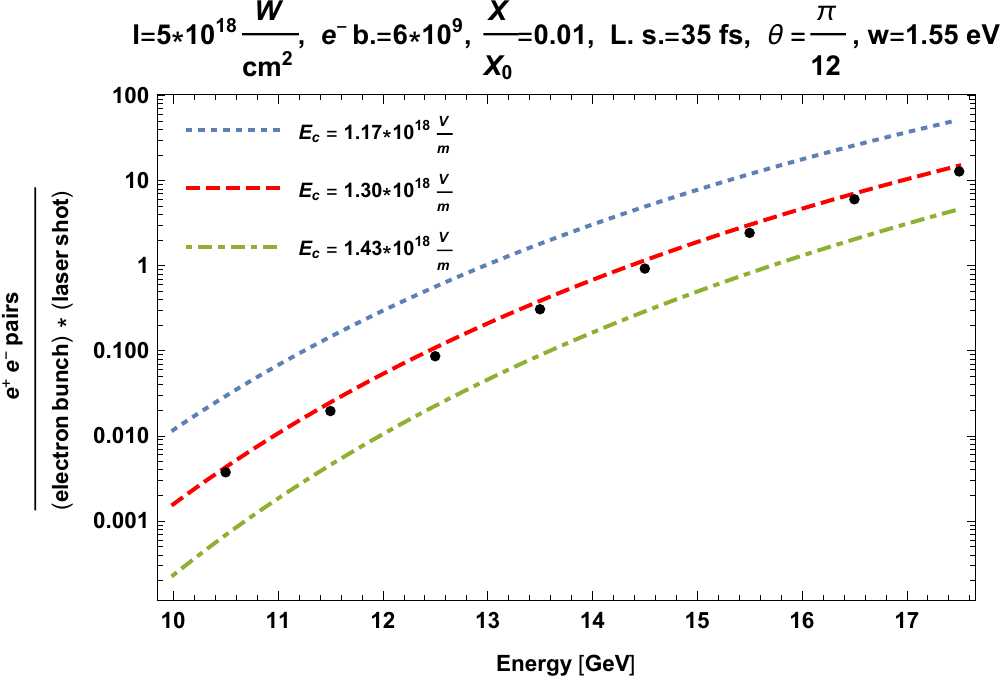}
\vspace{-2ex}
\caption{\AR{Total number of $e^+ e^-$ pairs produced per electron bunch ($6\times 10^9$ electrons of energy $E_e$) impinging on the bremsstrahlung target (thickness $X/X_0=0.01$) and per laser shot (duration $35\,\text{fs}$, laser frequency $\omega = 1.55\,\text{eV}$, intensity $I= 5\times 10^{18}\,\text{W}/\text{cm}^2$) 
crossed with the bremsstrahlung photons at an angle of $\theta = \pi/12$, as a function of $E_e$. 
The black dots exploit the full numerical result for specific values of the electron beam energy. 
The dashed line shows the
analytic prediction resulting from \eqref{pair_prod_circularly_pol_laser_large_intensity_thin_target_brems},
exploiting the relations \eqref{xi_intensity} and \eqref{chie_intensity}. The dotted (dot-dashed) line shows the same analytic prediction,
but for the case where the value of the Schwinger critical field ${\rm E}_c$ deviates by a multiplicative factor of $\kappa =0.9$ ($\kappa = 1.1$) from its nominal value 
\eqref{schwinger-crit}.
\label{fig:log_log_decay_rate_thin_target_brems_intensity_Ee}
}
}
\end{center}
\end{figure}

For the BPPP process, high energy electrons will impinge in bunches onto the target. The electron beam of the European XFEL, for example,  
contains $6\times 10^9$ electrons of energy \AR{up to} $E_e=17.5\,\text{GeV}$, with small energy spread and a good emittance \cite{Altarelli:2006zza}.  The high intensities of the laser are reached conceivably 
in laser pulses of duration around $35\,\text{fs}$, as in the LUXE experiment which is currently in the design phase for  
a proposal to the European XFEL  Facility \cite{luxe}. In Fig. \ref{fig:log_log_decay_rate_thin_target_brems}, we show 
the number of pairs produced per electron bunch and per laser shot expected in this case. The solid lines are obtained from 
the numerical solution of  Eqs. \eqref{eq:Foppp} and \eqref{electron_initated_laser_assisted_OPPP},
while the dashed lines exploit the analytic asymptotics \eqref{pair_prod_circularly_pol_laser_large_intensity_thin_target_brems}.
Importantly, the latter approaches the former already at $\xi\gtrsim 1$ and $\chi_e\lesssim 1$. 
Moreover, the number of produced pairs is favorably high, even for the most interesting parameter range of large $\xi$ and small $\chi_e$. 
From this we conclude that this type of experiment allows for  a determination of the Schwinger critical field ${\rm E}_c$
by fitting the experimentally determined number of produced $e^+e^-$ pairs to the theoretical prediction,  \eqref{pair_prod_circularly_pol_laser_large_intensity_thin_target_brems} or, equivalently, \eqref{eq:SPP_BPPP}, using ${\rm E}_c$ as a fit parameter, cf. Figs. \ref{fig:log_log_decay_rate_thin_target_brems_intensity} (top panel) and \ref{fig:log_log_decay_rate_thin_target_brems_intensity_Ee}. 

In practice, in an experiment as sketched in Fig. \ref{fig:experiment_sketch}, it will be easiest to change the intensity of the laser and 
the energy $E_e$ of the electron 
beam. In this case, the electron recoil parameter can be expressed as
\AR{\begin{equation}
\label{chie_intensity}
\chi_e = 0.1576  \left( 1 + \cos\theta \right)  \left( \frac{E_e}{17.5\,\text{GeV}}\right) 
\left( \frac{I}{10^{19}\,\text{W}/\text{cm}^2}\right)^{1/2} \,.
\end{equation}}
The predicted number of electron pairs per electron bunch and per laser shot are presented in Fig. \ref{fig:log_log_decay_rate_thin_target_brems_intensity}, for 
fixed electron beam energy, as a function of the laser intensity, and in Fig.  \ref{fig:log_log_decay_rate_thin_target_brems_intensity_Ee}, for fixed laser intensity, as a function of the 
electron beam energy. 
We infer that the number of produced pairs per electron bunch and laser shot rapidly grows with increasing intensity $I$ and electron beam energy $E_e$ to values above 1 already at  \AR{$I\sim 5\times 10^{18}$\,W/cm$^2$} and  $E_e\sim 14$\,GeV.
Therefore, the asymptotic regime for the BPPP process should be experimentally accessible with reasonable accuracy at the 
European XFEL faciltiy, requiring only 
 modest parameters for a focussed intense laser to ensure stable operation at a strong field experimental interaction point. 
This will allow a precision comparison with the asymptotic result according to 
Eq. \eqref{pair_prod_circularly_pol_laser_large_intensity_thin_target_brems}, which sensitively depends on the value of ${\rm E}_c$, cf.  
Figs.  \ref{fig:log_log_decay_rate_thin_target_brems_intensity} (top panel) 
and \ref{fig:log_log_decay_rate_thin_target_brems_intensity_Ee}: A variation of ${\rm E}_c$ around its nominal value 
\eqref{schwinger-crit} by 10\,\% results
in a change in the predicted rate by nearly an order of magnitude.
By removing the target in the experimental setup of Fig. \ref{fig:experiment_sketch}, the strong-field trident process can be studied in addition. In its two-step variant, it occurs via high intensity Compton scattering (HICS), followed by OPPP. Exploiting in this way the energetic photons from HICS as an alternative source of high energy photons, the asymptotic regime of the OPPP process can again be in principle measured. However, the rate for HICS is considerably lower and is cut off by the stepped Compton edge, compared to that of bremsstrahlung. Nevertheless, trident pair production is of significant interest \cite{Bula:1997eh,Hu:2010ye,Ilderton:2010wr,King:2013osa,Dinu:2017uoj} since its measurement in the 1990s by the SLAC E144 experiment exploiting the 46.6 GeV electron beam\footnote{At this energy, multiple laser photons are required in order to produce the OPPP pair. Pure photon-photon pair production could only be achieved with a 200 GeV electron beam. In the absence of such higher energy electron beams, an alternative way to achieve sufficient centre of mass energy to produce a pair may be via bremsstrahlung photon interaction inside a laser-heated blackbody radiation cavity \cite{Pike:2014wha}. A similar idea, but with a different scheme to reach the pair production threshold via high energy photons, is proposed by \cite{Ribeyre:2015fta}. In this case, two interacting beams of MeV level photons are produced by impinging lasers on solid or gas targets. Though these schemes potentially produce copious amounts of pairs, the asymptotic regime of interest in this paper, $\xi\gtrsim 1$, does not appear to be reachable in the foreseeable future.} of the Stanford Linear Accelerator \cite{Bamber:1999zt} and is an important additional strong field process that can be measured by the experiment described in this paper.

In order to quantify the expected experimental accuracy, detailed simulations will have to be carried out. These will include a GEANT 
 \cite{Agostinelli:2002hh} model of the converter target to produce a large flux of energetic photons. Macro-particles, representing a train of electron bunches varying randomly within known beam conditions, will also be required. An accurate representation of laser pulse shape and jitter will be needed, as well as a full accounting for crossing angles and beam overlap. Along with GEANT, a full strong field QED 
particle-in-cell code including higher order processes has also to be developed.

\subsection{Effect of finite laser pulse length}

A substantial amount of analytical work on strong field processes over the last decade has concerned itself with the fact that a real experiment will utilise a real laser pulse. The infinite plane wave approximation (IPW) which we have exploited in order to determine the OPPP transition rate neglects the finite length and shape of the laser pulse.  Allowance for these realistic features  has the potential to significantly complicate the transition rate expression. Nevertheless, various additional approximations have been employed in order to produce manageable results.

If the pulse length is not too short in comparison to the laser wavelength, the slowly varying envelope approximation (SVEA) can be employed. This requires the assumption that the pulse envelope does not change much over the course of one wavelength of the laser field. For LUXE, an 800 nm laser with a 35 fs pulse length is envisaged \cite{luxe}. This gives $N\mathcal\approx 12$ periods within the pulse, meaning that the SVEA may be a reasonable assumption.

The existing SVEA calculations can be utilised with the aim of testing the asymptotic limits of the OPPP rate. Returning to the matrix element of the OPPP process, the IPW describes a definite number of laser photons participating in the process expressed by a delta function. A finite pulse however smears out the longitudinal light cone (four-vector, $z\equiv(\vec{z}_{+},z_{-})$) momentum transfer and the delta function is replaced with a pulse \cite{Heinzl:2010vg},
\begin{eqnarray}
\nonumber
S_{fi}&=&\sum_n \delta^{(4)}\!\left(z\right)\, M_n\quad ,z\equiv p+q-k_i,\hfill\quad\text{(IPW)}, \\
\nonumber
S_{fi}&=&\sum_n \delta^{(3)}\!\left(\vec{z}_{+}\right)\frac{\sin N\pi z_{-}\sinc\pi(z_{-}-n)}{\sin \pi z_{-}}\, M_n,\hfill\quad\text{(SVEA)}.
\end{eqnarray}

For a square pulse, there is sub-threshold as well as envelope behaviour. However, the delta comb structure is largely restored as the pulse length increases to $N=4$, suggesting that the IPW approximation is satisfactory for the experimentally planned pulse length of $N\approx 12$. To be sure of the asymptotic behaviour of the OPPP process though, a full calculation of the transition probability with a realistic pulse shape is necessary.

Such a calculation has been performed and analysed for a range of intensities and pulse lengths
in Refs. \cite{Nousch:2012xe,Titov:2012rd,Titov:2013kya,Titov:2015pre}.
In the range of very high and very low intensity parameter $\xi$ the calculation of the OPPP with finite pulse length was performed 
analytically. In the intermediate range, the calculation was carried out numerically,  across a range of photon energies and incident angles. The IPW was shown to be a very good assumption by the time the pulse length increased to $N=10$, meaning that the IPW analysis performed in this paper is suitable for a LUXE type experiment.

\subsection{Effect of variable laser intensity}

In a real strong field experiment, with a laser pulse focussed to a small spot size, the infinite plane wave assumption can also be questioned in regards to directions perpendicular to the pulse propagation. One approach is to develop electron wave functions embedded explicitly in an external field 4-potential with an appropriate transverse description. The equations of motion can be solved approximately, by assuming that the electron energy is the largest dynamical parameter, which is reasonable for experiments with ultra relativistic electrons \cite{DiPiazza:2015xva}.

The solutions which allow for the transverse focussing lead to the conclusion that the Volkov solution with IPW assumption can be used as long as the local value of the strong laser intensity, at the point of pair production, is taken into account. This is necessarily the case for any strong field PIC code that simulates the pair production in a real bunch/pulse interaction \cite{Hartin:2018egj}.
 
The use of a strong field PIC code in future experimental studies will allow the full variation of the laser pulse, both longitudinally as well as transversely, to be taken into account. Experimentally, the high intensity laser pulse is expected to have a flat intensity in the transverse direction, but a Gaussian shape longitudinally. A real experiment will seek to account for the Gaussian shape with some sort of analysis filter or by making kinematic cuts to detector signals. 

Another experimental possibility is to deliberately form the pulse length so that the laser intensity is relatively constant throughout the interactions of interest. Since there is a relationship of the laser intensity $I$ to pulse energy $E$ as well as pulse length $\tau$ and pulse width $w_0$, it is possible to adjust the former in order to set the latter to desired values,
\begin{equation}
I=\frac{E}{\pi w_0^2\tau}.
\end{equation}

A further, extensive experimental study will be necessary in order to determine suitable operating conditions. Such a study will take into account the angular spread of bremsstrahlung from the foil, the distance between foil and strong laser interaction point, and the characteristics of the detector system recording the resultant produced pairs.

\section{Conclusions and Outlook}
\label{sec:conclusions}

The theory study and experiment, presented here, already promises us the first ever measurement of the Schwinger critical field value, through an asymptotic limit. Many additional, important strong field effects can be experimentally tested through further theoretical and phenomenological studies.

One may ask what happens if the quantum recoil parameters continue to increase in value with either increasing gamma energy and/or electromagnetic field intensity \cite{Fedotov:2010ja}. In such a case, our strong field experiment would probe both smaller distances and the quantum vacuum would be increasingly polarised. In such circumstances, there is good reason to think that higher order, strong field processes may increasingly play a role. 
There is not only the opportunity to study and search for such higher order processes in dedicated strong field experiments, there is also the ability to question non-perturbative quantum field theory itself. 
The Furry picture \cite{Furry:1951zz}, which is the semi non-perturbative theory from which most strong field QED predictions are made, includes the quantum recoil parameters in its effective coupling constant. Theoretical estimates \cite{Narozhnyi:1979at,Ritus:1985} put the effective coupling constant at $\alpha\chi^{1/3}$, meaning that the theory breaks down at a high enough value of the recoil parameters. Whereas this breakdown regime is experimentally some distance away, the value of the effective coupling constant through the higher order terms is possibly in reach. \\

\mysections{Acknowledgments}

A.R. would like to thank Paola Arias, Holger Gies, Axel Lindner, Gerhard Paulus, Javier Redondo, and Andreas Wipf 
for early discussions on the possibility of 
experiments exploiting the electron beam of the European XFEL and a high-intensity laser to measure non-linear and non-perturbative processes predicted by QED. 
We would like to thank Beate Heinemann for a careful reading of the manuscript and useful suggestions.
A.H. acknowledges support from a Leverhulme Trust Research Project Grant RPG-2017-143 and by STFC, United Kingdom.
N.T. thanks to the Conicyt scholarship 21160064 and to the University of Santiago de Chile.\\



\end{document}